\documentstyle[sprocl,epsfig]{article}

\def\Journal#1#2#3#4{{#1} {\bf #2}, #3 (#4)}

\def\NPA{{\em Nucl. Phys.} A}
\def\PLB{{\em Phys. Lett.}  B}

\def\PRL{\em Phys. Rev. Lett.}
\def\PRC{{\em Phys. Rev.} C}
\def\PRD{{\em Phys. Rev.} D}

\def\be{\begin{equation}}
\def\ee{\end{equation}}
\def\bea{\begin{eqnarray}}
\def\eea{\end{eqnarray}}
\begin{document}

\title {   \bf  $\eta$ Production in Hadronic Interactions }
\author {Alfred \v{S}varc, Sa\v{s}a Ceci}

\address{    Rudjer Bo\v{s}kovi\'{c} Institute, Bijeni\v{c}ka cesta 54, \\
             10000 Zagreb, Croatia \\
            E-mail: svarc@rudjer.irb.hr, ceci@rudjer.irb.hr
                                                   }

\maketitle\abstracts
{A short, and definitely not a complete representation of $\eta$ production processes on hadrons is %%@
given. First of all, the different ways of obtaining the $\pi N \rightarrow  \eta N$ and $\pi N %%@
\rightarrow \eta N$ amplitudes are presented. After that, an overview of results obtained using %%@
these amplitudes as input for calculating processes like: $NN \rightarrow \eta NN$ , $pd %%@
\rightarrow \eta ^{3}He$, $\pi d \rightarrow \eta NN$ and $\eta d$, $\eta ^{3}He$ and $\eta ^{4}He$ %%@
as well as $\eta$-light nuclei bond states, will be given. The experimental and theoretical results %%@
will be reviewed. The opened problems and the way how to solve
them will be presented.}

\section{Introduction}
           The problem of determining whether there exists a resonance in the N$^*$ system is a very %%@
nontrivial one. In addition to the problem of the definition what an N$^*$ resonance %%@
is\footnote{See the general discussion on the B(arion)R(esonance)A(nalysis)G(roup)-BRAG workshop %%@
preceding this workshop} there is a problem of coupling different resonances to the different %%@
channels. For example, S$_{11}$(1650) MeV resonance does not at all couple to the photoproduction %%@
channel, so it is practically invisible in all processes involving $\eta$ photoproduction. The %%@
second example is the fourth P$_{11}$ resonance, which is not visible in any process which does not %%@
involve $\eta$ production in hadronic reactions. Therefore, looking at photoproduction processes %%@
only it is not sufficient to see all possible N$^*$  resonances. Consequently, the only method %%@
which can be applied selfconsistently to obtain all resonances in all channels is a multichannel, %%@
multiresonance, unitary coupled channel model developed by \cite{Cut79}, and maximizing and %%@
updating the input to the model.
In order to value the strength of the used method, the amount of the worldwide work involved in the %%@
analysis, the mutual agreement of the results and the competence of the authors, we have decided to %%@
rank the publications by a number of stars, very similar to the method used by Particle Data Group %%@
(PDG). One star indicates the pioneering attempts, while four star denotes the general world %%@
interest and a significant level of agreement reached. The estimate is just a personal judgement of %%@
the authors of the article and anyone is welcomed to modify
 it.

\section { $\pi N \rightarrow \eta N$ and $\eta N \rightarrow
\eta N$ models  }

\subsection { Coupled channel  models  $^{****}$ }

Multiresonance, coupled channel and unitary models offer the best possibility to treat all the %%@
channels simultaneously, without the problem of not seeing the resonances which couple poorly to %%@
one of the channels. The framework has been elaborated by \cite{Cut79}, and has been used by most %%@
of the modern approaches. It is essential to remind the reader to use the original article by %%@
Cutkosky \cite{Cut79} where the full formalism is explained. The results of the original model %%@
insignificantly differ from the predictions of KH80 group \cite{Hoe83}, and represent the state of %%@
the art of knowledge of 80es in $\pi$-N physics. Later on, in 90es the chain of articles started to %%@
use the same formalism and exploit the new data to make more precise  conclusions %%@
\cite{Man92,Bat98,Vra99}.

 \begin{figure}
\begin{center}
 \epsfig{figure=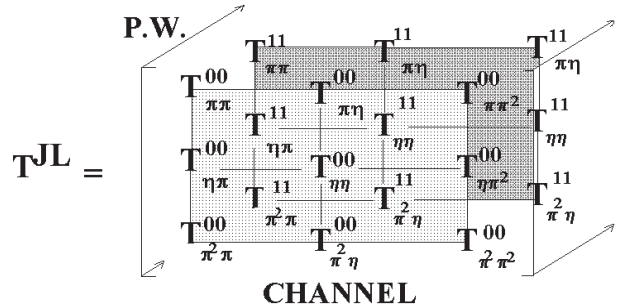,height=4cm,width=8cm}
 \caption{The 3-D T-matrix in
 Cutkosky formalism.
 \label{fig:1}}
\end{center}
 \end{figure}

The idea of all three approaches was basically the same: to use the well known and tested formalism %%@
\cite{Cut79}, and to introduce the new knowledge about $\eta$ production processes in order to %%@
obtain more reliable information about N$^*$ resonances. The essence of the formalism was to obtain %%@
the "three dimensional" T-matrix shown in Fig.1 for three channels and has been used in %%@
\cite{Bat98}. In that case the chosen channels ($\pi-N$, $\eta-N$ and the third effective two body %%@
channel $\pi^2-N$ are given on the x-axes, while partial waves are given on the y-axes). The whole %%@
formalism allows the separation of the T-matrix in partial wave amplitudes which are indicated by %%@
vertical planes in the 3-D T-matrix. However, the main problem in this formalism was a numerical %%@
minimization procedure which tended to explode in number of fitting parameters for bigger number of %%@
channels if experimental observables were fitted, because all partial waves are automatically %%@
mixed. Each of  three references have tried to overcome that problem in a different way. In refs. %%@
\cite{Man92,Bat98} authors have used three coupled channels only, but have chosen to fit the $\pi-%%@
N$ elastic T-matrices, and experimental observables. In  ref \cite{Man92} the $\pi$-N elastic %%@
partial wave T-matrices from different sources \cite{Cut79,Hoe83} have been used for the first %%@
channel, the second channel-$\eta N$  channel have only been represented by the  S$_{11}$(1535) %%@
resonance, and the whole known data set for continuum pion production $\pi N \rightarrow \pi \pi N$ %%@
have been used to represent the third channel. The number of parameters was acceptable, and the %%@
minimization has revealed results very similar to PDG group, but more constrained in other but $\pi %%@
N$ elastic channel. The second PWA \cite{Bat98} has as well used the $\pi-N$ elastic T-matrices %%@
from the same sources as \cite{Man92}, but have chosen to use the whole set of measured total and %%@
differential cross sections for the $\pi N \rightarrow \eta N$ process.

 In both cases the number of parameters to be fitted was quite big (of the order of 100), but the %%@
minimization procedure was still under full control. However, the drawback of both of these %%@
approaches was that it was not forseable to increase the number of coupled channels because the %%@
number of fitting parameters would explode beyond control if one uses MINUIT program. The third %%@
approach \cite{Vra99} has avoided the problems in that program by not fitting the experimental data %%@
but T-matrices obtained from different sources. In that way they have been able to fit partial wave %%@
by partial wave (or plane by plane in Fig.1), and that has significantly reduced the number of %%@
input parameters and allowed them to use much more then three channels. However, the choice of %%@
input T-matrices remains an opened question to be discussed and tested. The all three analysis show %%@
a fair level of agreement and self consistence, and we dare to say that T-matrices for $\pi-N$ and %%@
$\eta-N$ channel are quite confidently determined and can be used as the input for the calculation %%@
of more complicated processes.

\subsection { Quark model and coupled channel model $^{****}$ }

In ref \cite{Cap99} the importance of multichannel approach has been illustrated. Namely, the %%@
possibility of existence of the fourth P$_{11}$ resonance has been reported in spite of the fact %%@
that it has not been seen in any previous single channel analysis, or even in one three coupled %%@
channel analysis which included the channel into which that resonance does not couple \cite{Man92}. %%@
First the existence of that resonance has been predicted in the quark model \cite{Cap98}, but at %%@
the same time it was seen in the three coupled channel analysis \cite{Bat98}.
The agreement of the findings of both, theoretical and phenomenological analysis are striking, so %%@
it is quite likely that the number of P$_{11}$ N$^*$  resonances should be increased to four. The %%@
result is a good example of theoretical prediction confirmed by the "experimental" partial wave  %%@
 analysis. \\

\begin{center}
{\normalsize
\begin{tabular}{|c|c@{}c@{}c@{}c@{}c|c@{}c@{}c@{}c@{}c|}
 \hline
\multicolumn{1}{|c}{\bf PDG} & \multicolumn{5}{|c}{{\bf Zagreb group}} &  \multicolumn{5}{|c|}{\bf %%@
Quark model}       \\
\hline
     States          & \multicolumn{5}{|c}{Four resonances}    &  \multicolumn{5}{|c|}{Five %%@
resonances}   \\
                  \hline
   &    Mass        &   Width        & $x_\pi$      & $x_{\eta} $     & $x_{\pi^2} $    &    Mass        %%@
&   Width        & $x_\pi$      & $x_{\eta} $     & $x_{\pi^2} $    \\
L$_{2I,2J}$${\rm (_{Mass/Width}^{x_{el}})}$ &  (MeV)       &   (MeV)        &  (\%)        &   (\%)          %%@
&   (\%)    &  (MeV)       &   (MeV)       &  (\%)        &   (\%)          &   (\%)            \\
 \hline
           S$_{11}(_{1535/120}^{38})$     &    1553   &    182     &    46     &      50      %%@
&    4    &  1460    &   645 &   34  &   66  &   0   \\
    S$_{11}(_{1650/180}^{61})$     &    1652   &    202     &    79     &      13      %%@
&    8    &  1535    &   315 &   47  &   39  &   14  \\
    S$_{11}(_{2090/95 \: }^{9})$   &    1812  &    405     &    32     &      22     %%@
&    46   &  1945    &   595 &   6   &   2   &   89  \\ \hline
    P$_{11}(_{1440/135}^{51})$     &    1439  &    437     &    62     &      0       %%@
&    38   &  1540    &   425 &   97  &   0   &   3   \\
    P$_{11}(_{1710/120}^{12})$     &    1729  &    180     &    22    &      6       %%@
&    72  &  1770    &   305 &   6   &   22  &   72  \\
    P$_{11}$                       &    1740  &    140     &    28    &      12      %%@
&    60  &  1880    &   155 &   5   &   18  &   76  \\
    P$_{11}$                       &    -         &    -           &    -         &      -          %%@
&     -      &  1975    &   45  &   8   &   0   &   92  \\
    P$_{11}(_{2100/200}^{   9})$   &    2157  &    355     &    16     &      83      %%@
&     1   &  2065    &   270 &   22  &   1   &   77  \\ \hline
    D$_{13}(_{1520/114}^{54})$     &    1522   &    132     &    55     &      0.1   %%@
&    45   &  1495    &   115 &   64  &   0   &   36  \\
    D$_{13}(_{1700/110}^{   8})$   &    1817  &    134     &     9     &      14      %%@
&    77   &  1625    &   815 &   4   &   0   &   96  \\
    D$_{13}(_{2080/265}^{   6})$   &    2048  &    529     &    17     &      8       %%@
&    75   &  1960    &   535 &   12  &   6   &   81  \\
    \hline
\end{tabular}
  }
\vspace{0.25cm}
\end{center}

{\em Table 1:} Resonance parameters of the phenomenological
\protect{\cite{Bat97a}} and the quark \protect{\cite{Cap99}}
  models. The states are defined by the latest values given by the
    PDG group \protect{\cite{PDG}} and other parameters are defined
     in the text. Errors can be find in original
     publications.

\subsection{ $\eta$N S-wave scattering length $^{****}$}

      Another example of importance of the $\eta$ production in hadronic channels for understanding %%@
the structure of N$^*$ resonances is the need of the existence of the second S$_{11}$(1650) %%@
resonance (the resonance which is extremely poorly coupled to the photoproduction channels) for the %%@
complete understanding of the $\eta$N S-wave scattering length. Namely, the problem of extremely %%@
poorly determined value of the real part of the $\eta$N S-wave scattering length has been known for %%@
years, and the limits have been 0.2 fm $\leq$ Real($a_{\eta N}$) $\leq$ 0.98 fm. That ambiguity was %%@
directly prohibiting the estimate of the likelihood of formation of the $\eta$-light nuclei bound %%@
states, because the existence of these bound states was directly correlated to the value of the %%@
real part of the $\eta$N scattering length as it can be seen in
Fig.2.

      The spread in possible values of the real part of the $\eta$N  S-wave scattering length is given in %%@
Fig.2, and can be easily understood. The problem has been extensively addressed in ref. %%@
\cite{Bat96}. As it has been explained, any single resonance model (containing only one resonance %%@
in the S-wave) with the addition of the quite reliably measured and remeasured slope of the $\pi^- %%@
p \rightarrow \eta n$ total cross section near threshold can only give the values of the real part %%@
of   the $\eta$N scattering length fixed below $\approx$ 0.4 fm. For any value bigger then that, %%@
the existence of the second S$_{11}$ resonance (1650 in addition to 1535) has to be assumed.
The ambiguity has finally been resolved by getting the overlapping  results from two calculations %%@
based on entirely different formalisms \cite{Bat98,Gre97}. It is indicative that both approaches %%@
{\bf have} to include the existence of the second S$_{11}$ resonance. As the both publications used %%@
completely different formalisms (Cutkosky formalism \cite{Bat98} and K-matrix formalism %%@
\cite{Gre97}), and results coincide within the error bars, we conclude that the real part of the %%@
$\eta$N S-wave scattering length is quite well determined now.  The existence of the second %%@
S$_{11}$ resonance for the overall understanding of the results given in Fig.2. is, henceforth,  %%@
established. Let us just remind the reader that the second S$_{11}$ resonance is not seen in %%@
photoproduction processes. On the basis of these results for the real part of the $\eta$ N S-wave %%@
scattering length the predictions for the existence of the bound states in different $\eta$ -light %%@
nuclei are given in Fig.2.

\noindent
\begin{figure}[h]
\label{fig:2}
\centerline{\epsfig{figure=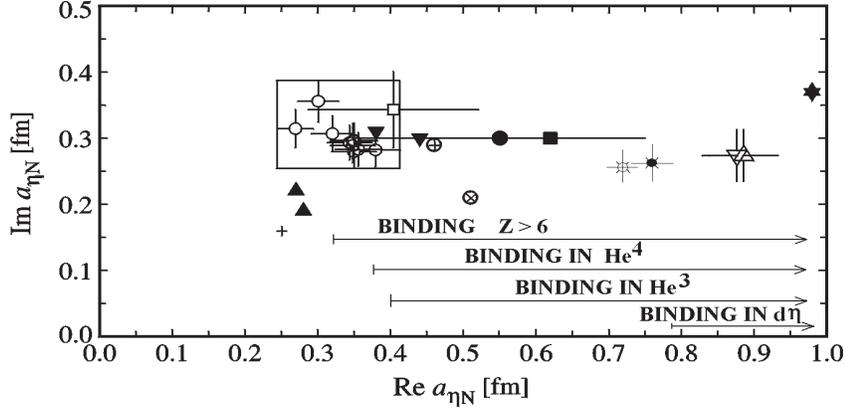,height=6cm,width=12.cm,angle=0}}
\caption{
$\eta$N S-wave scattering length. The symbols for all extracted values of the $\eta$N scattering %%@
length are taken over from reference \protect \cite{Bat96}. The only addition are the values %%@
extracted in ref. \protect \cite{Bat98} - crossed empty circles and \protect \cite{Gre97}  - %%@
crossed full circles. Lines given on the figure indicate for which values there is a probability %%@
for the $\eta$-light nuclei bound states \protect \cite{Wyc96}  }
\end{figure}

\subsection{ Other  $\pi N \rightarrow \eta N$ and $\eta N \rightarrow \eta N$ models $^{***}$ }

 Numerous other models have been produced with the aim to extract PW T-matrices for $\eta$ meson %%@
production in hadronic reactions \cite{Bha85,Ben91,Ari92,Aba96,Den98}. All of them suffer from some %%@
of the drawbacks: they are either single resonance, or simplified in some way. However, they are %%@
valuable to be looked at in order to see other approaches and possible simplification valuable for %%@
some specific cases.

\section { $NN \rightarrow NN \eta$ processes $^{**}$ }

     The knowledge of elementary PW $\pi N$ and $\eta$N amplitudes have been tested in calculations %%@
involving more then  two bodies in order to test the reliability and self consistence of obtained %%@
partial waves. One of the simplest examples is the $\eta$ production in nucleon-nucleon scattering. %%@
The process has been investigated experimentally \cite{Ber93,Chi94,Cal98,Cal99} and theoretically %%@
\cite{Bat97a,Vet91,Lag91,Wil93,Fal96a,Ber98}. However, even the initial agreement in theoretical %%@
calculations which mechanism is dominating has not been reached. It is generally agreed that in %%@
addition to the Born term the final state interaction should be added, but the combination of %%@
exchanged mesons which are described in different models varies. Therefore, more theoretical and %%@
experimental effort should be done in order to bring the problem to the general agreement.

\section {$\pi d \rightarrow NN \eta$ processes $^{**}$ }

    The testing of elementary PW $\pi$N and $\eta$N amplitudes is as well attempted for $\eta$ %%@
production processes in $\pi$d reaction. The experiments are scarce \cite{NefE890}, and theoretical %%@
calculations are just being developed \cite{Bat97b,Gar99,Gar00}. Among reproducing the various %%@
experimental quantities like total and differential cross sections, the idea of extracting the
$\pi^0$-$\eta$ mixing angle using ratios of $\pi^- d \rightarrow \eta nn$ and $\pi^+ d \rightarrow %%@
\eta pp$ has been suggested \cite{Bat97b}. However, the experimental analysis \cite{NefE890} has %%@
not yet been finished, therefore the comparison with the theoretical predictions is in a way %%@
"hanging in the air".

\section { Bound states of $\eta$ mesons $^{**}$ }

The attempts of finding indications of bound states in $\eta$-$^3$He system %%@
\cite{Fal95,Fal96b,Wil97} have been done. Results are, according to my belief, still opened to %%@
reader's interpretation.

The same statement stands for finding $\eta$-$^4$He bound states what has been attempted in refs. %%@
\cite{Wil93,Wil97,Fra94,Wil94,Cec99}, as well for finding $\eta$-light nuclei bound states in refs.
\cite{Ben90,Fix97,Kul98}.

As a final conclusion we tend to offer the \vspace*{0.5cm} statement:\\
   {\bf The existence of any N$^*$  resonance have to be confirmed in all channels, therefore, %%@
coupled channel models offer the best possibility to establish
them unambiguously.}

\end{document}